\documentclass[12pt]{article}
\usepackage{graphicx, epsfig, subfigure}
\usepackage[normal]{caption}
\newcommand{\beq}{\begin{eqnarray}}
\newcommand{\eeq}{\end{eqnarray}}
\newcommand{\baq}{\begin{eqnarray*}}
\newcommand{\eaq}{\end{eqnarray*}}

\newcommand{\rw}{\rightarrow}
\newcommand{\pr}{\prime}
\newcommand{\grad}{\bigtriangledown}

\begin{document}

\title{Charge-conjugation violating neutrino interactions in supernovae}
\author{
C.J.Horowitz \footnote{email: charlie@incf.indiana.edu}
 and Gang Li\footnote{email: ganli@indiana.edu } \\
Dept. of Physics and Nuclear Theory Center\\
Indiana University\\
Bloomington, IN 47405
}

\date{\today}

\maketitle


\begin{abstract} 
The well known charge conjugation violating interactions in the Standard
Model increase neutrino- and decrease anti-neutrino- nucleon cross sections.
This impacts neutrino transport in core collapse supernovae through ``recoil"
corrections of order the neutrino energy $k$ over the nucleon mass $M$.
All $k/M$ corrections to neutrino transport deep inside a protoneutron
star are calculated from angular integrals of the Boltzmann equation.
We find these corrections significantly modify neutrino currents at high 
temperatures.  This produces a large mu and tau number for the protoneutron
star and can change the ratio of neutrons to protons.  In addition,
the relative size of neutrino mean free paths changes.   At high temperatures,
the electron anti-neutrino mean free path becomes {\it longer} than that for 
mu or tau neutrinos.
\end{abstract}

Core collapse supernovae are perhaps the only present day large systems
dominated by the weak interaction.  They are so dense that only
neutrinos are known to diffuse.  This may allow the study of macroscopic
manifestations of the symmetries and features of the weak interaction.

We believe it is useful to try and relate some supernova properties to
features of the standard model.  Microscopic weak interaction
cross sections may influence macroscopic properties.  For example,
parity violation in a strong magnetic field may lead to an asymmetry
in the explosion and a recoil of the neutron star\cite{parity}.

Neutrino transport in dense matter is of great interest in astrophysics
and has been investigated by many authors \cite{T+S,B+P,S+S,B+V,
Iwamoto,Schinder}.  Some simulations of the cooling of proto-neutron stars 
based on these works can be found in References \cite{B+L,K+J}.
However, most previous work did not include the well known charge conjugation
violating interactions of neutrinos in the standard model.  These
interactions decrease neutrino currents and increase
anti-neutrino currents and may have many implications for supernovae.

For neutral currents, C violation leads to a large mu and tau number
for the proto-neutron star\cite{munumber}.  Anti-neutrinos have a longer
mean free path in matter than neutrinos.  Therefore, even if $\bar\nu_\mu$
and $\nu_\mu$ are produced in pairs, the anti-neutrinos escape faster leaving 
the star neutrino rich.  The mu and tau number may rise as high as $10^{54}$.
Supernovae may be the only known systems with large mu and or tau number.

For charged currents, C violation can change the equilibrium ratio
of neutrons to protons.  Neutrino capture on neutrons is favored over
anti-neutrino capture on protons.  This reduces the number of neutrons
in the neutrino driven wind above a proto-neutron star by 20 \% and
may significantly hinder r-process nucleosynthesis\cite{H+L3}.

Supernovae are complicated.  Therefore it is important to take
advantage of insight from symmetries.  Charge conjugation 
symmetry relates particles and anti-particles.  The interaction in 
Quantum Electrodynamics conserves C symmetry.  As a result, 
the cross section for $e^-p$ scattering is equal to that for $e^+p$ 
scattering, to lowest order in alpha.  In contrast, the cross
sections for $\nu_e$-nucleon  scattering and capture  are systematicly
larger than those for $\bar\nu_e$-nucleon.  
This difference comes from a number of sources.  First, isospin violation
in the nucleon leads to a mass difference $\Delta M$ between the neutron and
proton.  This will increase the neutrino capture cross section and
decrease the anti-neutrino capture cross section by a term of order $\Delta
M/k$ for neutrino energy $k$.  There are also coulomb effects of order 
$\alpha$.  Finally, there are terms coming from the $C$ violating nature
of the standard model weak interactions.  As we will see below, these
increase neutrino, relative to anti-neutrino, cross sections by of 
order $k/M$ for nucleon mass $M$.

Charge conjugation is closely related to parity P symmetry.  The product
CP is approximately conserved and P is maximally violated.  Therefore
there is large C violation.  However, time reversal symmetry limits the
effects of C violation for low neutrino energies.  Time reversal symmetry 
can relate neutrino-nucleon scattering to anti-neutrino-nucleon scattering 
where the nucleon scatters from final momentum $p^\pr$ to initial momentum 
$p$.  Thus the neutrino and anti-neutrino cross sections will be equal if the
nucleon does not recoil $p^\pr\approx p$.  Therefore we expect C violating
effects to be of recoil order $k/M$.  However, the recoil correction involves
the large magnetic moment of the nucleon, see below.

In this paper we calculate neutrino and anti-neutrino currents
in dense matter including all terms of order $k/M$ or equivalently
$T/M$ where $T$ is the temperature.  We focus on differences between
neutrino and anti-neutrino currents from the C violating interactions.
Note, there have been some relativistic calculations which implicitly
include the differences between $\nu$ and $\bar\nu$ interactions to all
orders in $k/M$.  See for example reference \cite{Reddy}.  However, this
calculation includes many effects.  By expanding to order $k/M$ we isolate
$\nu$ and $\bar\nu$ differences in a reasonably simple way.

The transport of neutrinos inside a supernova core is complicated. 
One needs to consider Pauli-blocking effects for nucleons and neutrinos.
Furthermore, during the cooling phase, electrons are
captured and the composition of the protoneutron star changes.
Finally, strong interactions between nucleons modify cross sections
\cite{H+W,Reddy}.  For simplicity this paper assumes non-degenerate, 
and noninteracting nucleons.  We will discuss the general case where 
nucleons can be degenerate in a later paper\cite{H+L5}.

Inside a supernova core, electrons are strongly degenerate and relativistic,
thus $\nu-e$ scattering is largely suppressed \cite{J+M}. Therefore  
we will consider neutrino-nucleon interactions only. For neutral currents, 
we have,
\beq
	\nu_i + n \rw \nu_i + n,  \\
	\nu_i + p \rw \nu_i + p,
\eeq
and for charged currents,
\beq 
        \nu_e + n \rw   e^- + p,   \label{eq:ch1}\\
 	\bar{\nu}_e + p \rw  e^+ + n. \label{eq:ch2}
\eeq

At neutrino energies of interest($\sim 10 $ Mev) the Standard 
Weinberg-Salam-Glashow \cite{W+S+G} model reduces to the Fermi theory 
with transition matrix element,
\beq
{\cal M}= \frac{G_F}{\sqrt{2}} 
[\bar{\nu}_i \gamma_{\mu}(1-\gamma_5) \nu_i][\bar{u}_j J^{\mu} u_j].
\eeq
In the above, $\nu_i$ and $u_j$ are spinors for neutrinos and nucleons
and $i=e,\mu,\tau$ and $ j=n,p$. The hadron current $J^{\mu}$ is, 
\beq
J^{\mu}=\gamma^{\mu}( c_v-c_a \gamma_5)+
        iF_2 \frac{ \sigma^{\mu \nu} q_\nu}{2M}. 
\eeq
Here $c_v$ is the vector coupling,  $c_a$ is the axial-vector coupling and 
$F_2$ describes weak magnetism, see Table I.  Note, the momentum dependence 
of the couplings from the finite size of the nucleon only enters at order 
$k^2/M^2$ and will be ignored.  We also ignore small coulomb effects and
terms proportional to the small electron mass.

\begin{table}
\centering Table I.  Coupling Constants ($g_a=1.26, \; \; {\rm sin} 
\theta_w^2 =0.23 $)
\medskip

\begin{tabular}{cccc}         \hline\hline \\
 reaction                        &  $c_v$                  &  $c_a$    & $F_2$       \\ \hline 
$\nu +n \leftrightarrow \nu +n $  &  $-1/2$                 &  $-g_a/2$ & $-0.972$    \\ 
$\nu +p \leftrightarrow \nu +p $  &  $1/2-2{\rm sin} \theta_w^2$  &  $g_a/2$  & $1.029$     \\
$\nu +n \leftrightarrow e^- + p $ &  $1/2$                  &  $g_a/2$  & $1.853$     \\ 
\hline \hline
\end{tabular}
\end{table}

The differential cross section for $\nu$ or $\bar\nu$ scattering from 
momentum $k$ to $k^\pr$ off of a nucleon of energy $E$ is,
\beq          
\frac{d\sigma}{d \Omega} = 
\frac{1 }{128 \pi^2} \frac{(k^{\pr})^2}{|d \epsilon / d k^{\pr}|}
 \frac{| \bar{\cal M}|^2}{E E^{\pr} k k^{\pr}} \label{eq:DCS}
\eeq
where $E^\pr$ is the final nucleon energy and $\epsilon=E+k-E^\pr-k^\pr$.
The reduced matrix element $| \bar{\cal M}|^2 $ is defined,
\beq
 | \bar{\cal M}|^2 =\sum_{s,s^{\pr}} |{\cal M}|^2
             =\frac{G_F^2}{2} {\cal L} _{\mu \nu} {\cal C} ^{\mu \nu}. 
\label{eq:Mfi2}
\eeq
Here $s$ and $s^{\pr}$ are the initial and final spin of the nucleon and 
${\cal L}_{\mu \nu}$ and ${\cal C}_{\mu \nu}$ are the lepton tensor and hadron 
tensor respectively, 
\beq
{\cal L}_{\mu \nu} &=& 8 \{ k^{\pr}_{\mu} k_{\nu} + k^{\pr}_{\nu} k_{\mu} 
                         - g_{\mu \nu} k^{\pr} \cdot k 
  \pm i \epsilon_{ \alpha \mu \beta \nu} k^{\pr \alpha} k ^{\beta} \},\\
{\cal C}_{\mu \nu} &=& {\rm tr}(\not{p}^{\pr}+m) J^{\mu} (\not{p} +m ) 
\bar{J^{\nu}},   \nonumber   \\
                  &=& {\rm tr} (\not{p}^{\pr}+m)
[  \gamma^{\mu} (c_v - c_a \gamma_5)  
  + \frac{iF_2 \sigma^{\mu \rho} q_{\rho}}{2M} 
     ]  \nonumber \\
& & \mbox{} (\not{p}+m)  
[   \gamma^{\nu} (c_v - c_a \gamma_5)
     - \frac{iF_2 \sigma^{\nu \delta} q_{\delta}}{2M} ],
     \nonumber          \\
 &=& {\cal C}_0^{\mu \nu} +F_2 {\cal C}_1^{\mu \nu}+ 
F_2^2 {\cal C}_2^{\mu \nu}.
\eeq
The momentum transfered to the nucleon is $q=k-k^\pr$ and the upper (lower)
sign is for neutrinos (anti-neutrinos).
We ignore the ${\cal C}_2^{\mu \nu}$ term because it is of order 
$\frac{k^2}{M^2}$.  For ${\cal C}_0$ and ${\cal C}_1$, we have,
\beq
{\cal C}_0^{\mu \nu} {\cal L}_{\mu \nu} = & 64 {
 (c_a^2 - c_v^2) M^2 (k \cdot k^{\pr} ) + 
 (c_a \mp c_v)^2 (k \cdot p^{\pr}) ( k^{\pr} \cdot p ) 
+(c_a \pm  c_v)^2 (k \cdot p) ( k^{\pr} \cdot p^{\pr} )} \\
F_2{\cal C}_1^{\mu \nu} {\cal L}_{\mu \nu} = &
\pm 128c_aF_2 ( p \cdot  k^{\pr} + p \cdot k ) (k \cdot k^{\pr})+
128c_vF_2(k \cdot k^{\pr})^2. 
\eeq
Schinder \cite{Schinder} only considered the 
${\cal C}_0^{\mu \nu} {\cal L}_{\mu \nu}$ term and showed it contains a 
$T/M$ term which many previous simulations on cooling of neutron stars have 
omitted.  Indeed, $F_2{\cal C}_1^{\mu \nu} {\cal L}_{\mu \nu}$ also
contributes to order $T/M$ and it has opposite sign for $\nu-N$ and 
$\bar{\nu}-N$ interactions.  This weak magnetism term is responsible for much
of the difference between the $\nu$ and $\bar{\nu}$ interactions.

The diffusion of neutrinos inside a supernova core obeys the Boltzmann 
Equation,
\beq
\frac{\partial f^{\nu}(\vec{k})}{\partial t} + \hat{k} \cdot 
\grad f^{\nu}(\vec{k})  =  
\frac{\partial Source}{\partial t}  
+\sum_{s,s^{\pr}}
{\int \frac{d^3\vec{k^{\pr}}} {(2\pi)^3} }
{\int \frac{d^3\vec{p^{\pr}}} {(2\pi)^3} }
{\int \frac{d^3\vec{p}}       {(2\pi)^3} }  
 \Gamma(f \leftrightarrow i) \nonumber  \\
 \{   
f^{\nu}(\vec{k^{\pr}})[1-f^{\nu}(\vec{k})]
f^{N}(\vec{p^{\pr}})[1-f^{N}(\vec{p})] 
- f^{\nu}(\vec{k})[1-f^{\nu}(\vec{k^{\pr}})]
f^{N}(\vec{p})[1-f^{N}(\vec{p^{\pr}})] 
 \}, \label{eq:Boltzmann1}
\eeq  

\beq
\Gamma(f\leftrightarrow i)=
\frac{|{\cal M}|^2(2\pi)^4 \delta^4(k+p-k^{\pr}-p^{\pr})}
{(2k)(2E)(2k^{\pr})(2E^{\pr})(|\vec{v_a}-\vec{v_b}|)}.
\eeq
The relative velocity of the colliding nucleon and neutrino is
$|\vec{v_a}-\vec{v_b}|=1$.  Nucleons are in good thermal equilibrium, so
we can use a Fermi-Dirac distribution for the nucleons $f^N(p)$. For
neutrinos, we expand the phase space distribution $f^{\nu}(k)$, 
\beq
 f^{\nu}(\vec k)= f^{\nu}_0(k) +3\hat{k}\cdot \vec{h}(k) + ..., 
\label{eq:expansion}
\eeq
about the equilibrium distribution $f^\nu_0=1/[1+{\rm exp}(k/T-\eta)]$.
The $\vec h$ term describes a nonzero neutrino current from gradients of the 
temperature $T$ and chemical potential $\mu$, with $\eta=\mu/T$.

We start by considering only neutral current interactions from
pure neutron matter.  Later a mixture of protons 
and neutrons is included.  We then consider pure charged current interactions 
and finally both charged and neutral currents.

For pure neutral currents, the source term in Eq. (\ref{eq:Boltzmann1}) is
zero and  we expect $\vec h=\vec h_0$ to have the form,
\beq
\vec{h}_0 = -\frac{f_0(1- f_0)}{ 3\rho_n \sigma_0 k^2 } 
\left[
\frac{k}{T^2} (1+\alpha_1\frac{k}{M}+\alpha_2\frac{T}{M})\vec\grad{T} 
  \pm (1+\beta_1\frac{k}{M}+\beta_2\frac{T}{M})\vec\grad{\eta} 
\right],       \label{eq:h01}
\eeq
with the plus sign for neutrino and the minus sign for anti-neutrino currents.
Equation (\ref{eq:h01}) includes corrections of order $k/M$ and $T/M$.  
The $T/M$ terms come from the thermal motion of
the nucleons while the $k/M$ terms come from recoil.  The coefficients
of these terms $\alpha_i$ and $\beta_i$ may be different if
the current arises from a temperature or chemical potential gradient.  
This corresponds to calculating a thermal conductivity 
(related to $\alpha_i$) or 
chemical diffusion coefficient (related to $\beta_i$).
When $\alpha_i=\beta_i=0$, the current reduces to Fick's law,
\beq
\vec{h}_0\rightarrow \vec{h}^0_0 = - \frac{1}{3\rho_N 
\sigma_0 k^2 } \vec\grad f_0. 
\eeq
Here $\sigma_0$ is the transport cross-section divided by
$k^2$,
\beq
\sigma_0 = \frac{1}{k^2} \int d\Omega \frac{d \sigma}{d \Omega}
(1-cos \theta)
=\frac{2 G_F^2}{3 \pi}(c_v^2+5 \: c_a^2) 
\eeq
and,
\beq 
\vec\grad{f_0} = f_0(1-f_0)(\frac{k}{T^2} \vec\grad T  \pm \vec\grad{\eta}).
\eeq
Using the expression for the differential cross section, Eq. (\ref{eq:DCS}) 
and assuming nucleons are non-degenerate, one gets from 
Eq. (\ref{eq:Boltzmann1}), 
\beq
\frac{\partial f}{\partial t} + \hat{k} \cdot \vec\grad f  =    
\frac{- f_0 (1-f_0) }{ \sigma_0} 
< \int  d \Omega_{ k^{\pr} }
\frac{ d \sigma }{ d \Omega_{k^{\pr}} } 
\frac{ k^{\pr 2} }{k^2} 
\Bigl\{ 
\frac{k}{T^2} (1+\alpha_1\frac{k}{M}+\alpha_2\frac{T}{M}) 
[(\frac{k}{k^{\pr}})\hat{k^{\pr}}- \hat{k} ]\cdot\vec\grad{T}
\nonumber \\
  \pm (1+\beta_1\frac{k}{M}+\beta_2\frac{T}{M})
[(\frac{k}{k^{\pr}})^2\hat{k^{\pr}}- \hat{k} ]\cdot\vec\grad{\eta}
\Bigr\} >_{\vec{p}}. 
\label{eq:Boltzmann2}
\eeq
Here the average of a quantity $A$ over the momentum of the initial 
nucleons is,
\beq
<A>_{\vec p}=\int d^3p f^N(p) A / \int d^3p f^N(p).
\eeq
The angular integral, $\int d\Omega_{\vec{p}}$,  can be done using
$<  \hat{p} \cdot {\vec{v}_1 }   \hat{p} \cdot {\vec{v}_2 }   
  >_{\Omega_{\hat{p}}} = \frac{1}{3}\vec{v}_1 \cdot \vec{v}_2$ 
for any vectors $\vec{v}_1$ and $\vec{v}_2$.   
The average over the magnitude of p is easy for non-degenerate nucleons,
$<p^2>_p = 3MT$.

Now, Eq. (\ref{eq:Boltzmann2}) becomes, 
\beq
\frac{\partial f}{\partial t} + \hat{k} \cdot \vec\grad f  =    
\frac{- G_F^2 f_0 }{ 4 \pi^2 \tilde{\sigma_0} } 
\int  d \Omega_{ k^{\pr} } \nonumber\\
\Bigl\{ 
(1+\alpha_1\frac{k}{M}+\alpha_2\frac{T}{M})\frac{k}{T^2} 
\bigl[ 
\hat{k^{\pr}} (A_1 + B_1 \frac{k}{M} + C_1 \frac{T}{M} )
-\hat{k} (A_0 + B_0 \frac{k}{M} + C_0 \frac{T}{M} ) 
 \bigr] \cdot\vec\grad{T}  \nonumber \\
\pm (1+\beta_1\frac{k}{M} + \beta_2\frac{T}{M}) 
\bigl[ 
\hat{k^{\pr}} (A_2 + B_2 \frac{k}{M} + C_2 \frac{T}{M} )
-\hat{k} (A_0 + B_0 \frac{k}{M} + C_0 \frac{T}{M} ) 
\bigr] \cdot\vec\grad{\eta}
\Bigr\}.  \label{eq:Boltzmann3}
\eeq
The coefficients of $A_i$, $B_i$ and $C_i$ are functions of $cos \theta
=\hat{k}\cdot\hat{k^\pr}$, with $\theta$ the neutrino 
scattering angle, and $c_v$, $c_a$ and $F_2$,
\beq
A_l &=& (c_a^2 - c_v^2)(1-cos \theta) + 2(c_a^2+c_v^2), \\
B_l &=& (1-cos \theta)[(l-3)A_l \pm 4c_a(c_v+F_2) ],      \\
C_l &=& [ l^2-5l+7 -(l^2-5l+6) cos \theta  ] A_l
     -3(c_a^2 - c_v^2)(1-cos \theta) + 2(c_a^2+c_v^2)cos \theta.
\eeq

The final step is to take the first angular moment of 
Eq. (\ref{eq:Boltzmann3}) by multiplying
both sides by the unit vector $\hat{k}$ and integrating $\int d\Omega_k/4\pi$.
The left hand side $LHS$ is 
$ \frac{ \partial \vec{h} }{ \partial t} + \frac{1}{3} \grad \!  f_0 $. 
If we ignore $ \frac{ \partial \vec{h} }{ \partial t} $, it is simply,
\beq
LHS = \frac{1}{3} \vec\grad\! f_0  = \frac{1}{3}f_0(1-f_0) 
( \frac{k \vec\grad \!T}{T^2} \pm \grad \! \vec\eta ),
\eeq
while the right hand side $RHS$ becomes, 
\beq
RHS = \frac{1}{3} f_0 (1-f_0) 
\Bigl\{ 
\frac{k}{T^2} (1+\alpha_1\frac{k}{M}+\alpha_2\frac{T}{M})
(1+\gamma_1\frac{k}{M}+\gamma_2\frac{T}{M} )\vec\grad{T} \nonumber \\
 \pm (1+\beta_1\frac{k}{M}+\beta_2\frac{T}{M})
(1+\kappa_1\frac{k}{M}+\kappa_2\frac{T}{M} )\vec\grad{\eta}
\Bigr\}.
\eeq
Here angular integrals of $A_i$, $B_i$ and $C_i$ give, 
\baq
\gamma_1 &=& \epsilon_0 \pm \delta + \Delta \! \epsilon_1, \;\;\;\;
\gamma_2 = \zeta_0 + \Delta \! \zeta_1,  \\
\kappa_1 &=& \epsilon_0 \pm \delta + \Delta \! \epsilon_2, \;\;\;\;
\kappa_2 = \zeta_0 + \Delta \! \zeta_2,  
\eaq
where by using $c_a$,$c_v$ and $F_2$ for the neutron, we have,
\beq
\delta &=& \frac{8 c_a(c_v+F_2) }{ c_v^2+5 c_a^2} = 3.32,  \label{eq:delta} \\
\epsilon_0 &=& -3 \frac{c_v^2+7c_a^2}{c_v^2+5 c_a^2} = -4.066, \;\;\;
\Delta \! \epsilon_1 = \frac{2 c_a^2}{c_v^2+5 c_a^2} = 0.355,  \;\;\;
\Delta \! \epsilon_2 = \frac{4 c_a^2}{c_v^2+5 c_a^2} = 0.71, 
\label{eq:epsilon}\\
\zeta_0 &=& \frac{40c_a^2+12 c_v^2}{ c_v^2+5 c_a^2} = 8.45,   \;\;\;\;
\Delta \! \zeta_1 = \frac{-8 c_a^2}{c_v^2+5 c_a^2} = -1.42,  \;\;\;\;
\Delta \! \zeta_2 = \frac{-12 c_a^2}{c_v^2+5 c_a^2} = -2.13.  \label{eq:zeta}
\eeq
Equating the LHS to the RHS, we can identify $\alpha_i$ and $\beta_i$ in
Eq. (\ref{eq:h01}) to be, 
\beq
\alpha_1 &=& - \gamma_1 = - (\epsilon_0 + \Delta \epsilon_1 \pm \delta )
                        = 3.71 \mp 3.32,  \label{eq:alpha1}                \\
\alpha_2 &=& - \gamma_2 = - (\zeta_0 + \Delta \zeta_1 )
                        = -7.03,                    \\
\beta_1  &=& - \kappa_1 = - (\epsilon_0 + \Delta \epsilon_2 \pm \delta ) 
                        = 3.36 \mp 3.32,                  \\
\beta_2  &=& - \kappa_2 = - (\zeta_0 + \Delta \zeta_2 )
                        = -6.32.\label{eq:beta2}
\eeq
This choice reproduces our original guess, Eq. (\ref{eq:h01}) for the
current.  The current has recoil and nucleon motion corrections of 
order $k/M$ and $T/M$ and we have evaluated the coefficients.


For $\nu-p$ scattering, the current $\vec{h}_0$ has the same form as 
Eq. (\ref{eq:h01}) with $\rho_n$  replaced by the proton density $\rho_p$ and,
\beq
(\sigma_0)_p= {(c_v^2 + 5 c_a^2)_p \over (c_v^2+5c_a^2)_n} 
\sigma_0 \;\;\;  = 0.89 \sigma_0,
\eeq
and $\alpha_i$, $\beta_i$ are recalculated using couplings $c_v$, $c_a$ and 
$F_2$ for protons, see Table I.  Equations 
(\ref{eq:alpha1}-\ref{eq:beta2}) become,
\beq
\alpha_1^p &=&  3.80 \mp 2.71,\\
\alpha_2^p &=& -6.40,\\
\beta_1^p &=& 3.40 \mp 2.71,\\
\beta_2^p &=& -5.60.
\eeq
Note, we have left an $n$ label off of $\alpha_i$ and $\beta_i$ in Eqs. 
(\ref{eq:alpha1}-\ref{eq:beta2}) for clarity.

For scattering from a mixture of protons and neutrons we can combine the 
two interactions.  Define,
\beq
r_0 = \frac{\rho_p (\sigma_0)_p}{ \rho_n \sigma_0 }\;\;  
    = 0.89 \frac{Y_p}{Y_n}.
\eeq
Then, 
\beq
\vec{h}_0 = - \frac{f_0(1-f_0)}
                   { 3k^2 \rho_n \sigma_0 (1+r_0) }
\left \{ 
\frac{k}{T^2} 
( 1 + \bar{\alpha}_1 \frac{k}{M}
    + \bar{ \alpha}_2 \frac{T}{M}  )\vec\grad{T}   
\pm 
( 1 + \bar{ \beta}_1 \frac{k}{M}
    + \bar{ \beta}_2 \frac{T}{M}  )\vec\grad{\eta}
\right \}.  \label{eq:hx}
\eeq
The average coefficients are,
\beq
\bar{\alpha_i} = {\alpha_i + r_0 \alpha_i^p\over 1 + r_0},\\
\bar{\beta_i}={\beta_i + r_0 \beta_i^p\over 1 + r_0},
\eeq
for $i=1,2$.  These average coefficients are close to those for
pure neutron matter.

We now calculate the number $J$ and energy $F$ currents for mu and
tau neutrinos,
\beq
\vec{J} = \int {d^3k\over (2\pi)^3} \vec{h}_0
        = -\frac{T}{6 \pi^2\rho_n \sigma_0(1+r_0)}
          \Bigl\{ 
\bigl[ L_1(\pm \eta) [1+ \bar\alpha_2 \frac{T}{M}]
                  + L_2(\pm \eta) \bar\alpha_1 \frac{T}{M} \bigr]
\frac{\vec\grad T}{T}\nonumber \\
\ \ \ \ \ \ \  \pm \bigl[ L_0(\pm \eta) [1+ \bar\beta_2  \frac{T}{M}]
                  + L_1(\pm \eta) \bar\beta_1  \frac{T}{M} \bigr]\vec\grad\eta 
\Bigr\}
\label{eq:J}
\eeq
\beq
\vec{F} = \int {d^3k\over (2\pi)^3} k\, \vec{h}_0
        =  -\frac{T^2}{6 \pi^2 \rho_n \sigma_0(1+r_0)}
\Bigl\{ \bigl[ L_2(\pm \eta) [1+ \bar\alpha_2 \frac{T}{M}]
                  + L_3(\pm \eta)  \bar\alpha_1 \frac{T}{M} \bigr]
\frac{\vec\grad T}{T}\nonumber \\
\ \ \ \ \ \ \ 
\pm \bigl[ L_1(\pm \eta) [ 1+ \bar\beta_2  \frac{T}{M} ]
                  + L_2(\pm \eta)  \bar\beta_1  \frac{T}{M} \bigr]
                  \vec\grad{\eta} 
\Bigr\} 
\label{eq:F}
\eeq
where $L_i(\eta)$ is defined by,
\beq
L_i(\eta) &=&\int_0^\infty dx x^i f_0(1-f_0),
\eeq
with $x=k/T$ and $f_0=(1+{\rm e}^{x-\eta})^{-1}$.
Equations (\ref{eq:J},\ref{eq:F}) are conventional diffusion currents with
order $T/M$ corrections.

We now consider charged current reactions corresponding to Eqs. (\ref{eq:ch1})
and (\ref{eq:ch2}).
We ignore the mass difference between neutrons and protons
so that the kinematics are the same for charged and neutral currents.
The Boltzmann Equation for $\nu_e$ and $e^-$ capture can be written, 
\beq
\frac{\partial f^{\nu}(\vec{k})}{\partial t} + \hat{k} \cdot 
\vec\grad f^{\nu}(\vec{k})  =  
\sum_{s,s^{\pr}}
{\int \frac{d^3\vec{k}^{\pr}} {(2\pi)^3} }
{\int \frac{d^3\vec{p}_{p} } {(2\pi)^3} }
{\int \frac{d^3\vec{p}_n}       {(2\pi)^3} } 
\Gamma(f \leftrightarrow i) \nonumber  \\
 \{   
f^{e^-}(\vec{k^{\pr}})[1-f^{\nu}(\vec{k})]
f^{p}(\vec{p_p})[1-f^{n}(\vec{p_n})] 
- f^{\nu}(\vec{k})[1-f^{e^-}(\vec{k^{\pr}})]
f^{n}(\vec{p_n})[1-f^{p}(\vec{p_p})] 
 \}. 
\eeq
For $\bar\nu_e$ and $e^+$ capture the Boltzmann equation has the same
form with $f^{\nu}$, $f^{e^-}$ and $f^n$ interchanged with $f^{\bar{\nu}}$,
$f^{e^{+}}$ and $f^{p}$ respectively.

Chemical equilibrium relates the chemical potentials $\mu_i$ for $i=e,p,\nu$ 
and $n$, $\mu_{\nu}= \mu_e + \mu_p - \mu_n $. This simplifies the 
Boltzmann Equation.  Using Fermi-Dirac distributions
for protons, neutrons and electrons and Eq. (\ref{eq:expansion}) for 
neutrinos, we find, 
\beq
 \frac{\partial f^\nu}{\partial t} + \hat{k} \cdot \vec\grad f^\nu  
= \frac{\rho_n}{128 \pi^2} 
< \! \int d \Omega_{k^{\pr}} 
 [1-f^{e^-}(\vec{k})] 
\frac{1}{d \epsilon /dk^{\pr} } 
\frac{k^{\pr 2} | \bar{\cal M} |^2  }{E E^{\pr} k k^{\pr} } 
 \{  \frac{-3 \vec{h}_{ch} \cdot \hat{k}}{1- f_0^{\nu}(k)}  \}
\! >_{\vec{p}}.
\eeq
Here, the factor of $1/(1-f_0^\nu)$ describes stimulated 
absorption\cite{I+N}.
The matrix element $|\bar{\cal M}|^2$ is given by Eq. (\ref{eq:Mfi2})
multiplied by an additional factor of $4{\rm cos}^2\theta_c$ with
$\theta_c$ the Cabbibo angle.
One finds the neutrino current to be, 
\beq
\vec{h}_{ch} = -\frac{R}{3 \rho_n \sigma_{ch}k^2 } 
\{
1+ \alpha_1^{ch}\frac{k}{M} + \alpha_2^{ch} \frac{T}{M} 
\}
\vec\grad{f^{\nu}_0(k)},   \label{eq:hch}
\eeq
where 
\beq
 \sigma_{ch} = 
 \frac{4 G_F^2 cos^2 \theta_c}{\pi}(3c_a^2 + c_v^2) 
\eeq
is the zeroth order charged current cross-section divided by $k^2$ and, 
\beq
R = \frac{1-f_0^{\nu}}{1-f^{e}}  = \frac{ 1/ (1+ e^{-(k \mp\mu_{\nu})/T}) } 
{ 1/ (1+ e^{-(k \mp \mu_{e})/T} ) } = 
\frac{ 1+ y^{\pm 1} e^{-(k \mp \mu_{\nu})/T} } { 1+ e^{-(k \mp \mu_{\nu})/T} }
=(1-f_0^{\nu}) + y^{\pm 1} f_0^{\nu}
\eeq
with
$$
y = e^{ - (\mu_{\nu}-\mu_{e})/T} = e^{ (\mu_{n}-\mu_{p})/T}
  = e^{ (\eta_{n}-\eta_{p})} =\frac{Y_n}{Y_p}
$$
As usual, the upper sign is for $\nu_e$ and the lower sign for $\bar{\nu_e}$.  
The coefficients in Eq.~(\ref{eq:hch}) are,
\beq 
\alpha_1^{ch} &=& 
\frac{2(c_v^2+5c_a^2)}{ 3c_a^2+c_v^2 } \mp 
\frac{4 c_a(c_v+F_2)}{ 3c_a^2+c_v^2 }
=3.10 \mp 4.12,  \\
\alpha_2^{ch} &=&
-\frac{4 (2 c_v^2+ 5 c_a^2)}{3c_a^2+c_v^2 } 
=-6.90.
\eeq
In general, R is k dependent. For simplicity we use a simple average which
allows us to illustrate the effects of the order $T/M$ corrections,
\beq 
R\rightarrow <R>= \int_0^\infty dk f_0^\nu(k) R / 
\int_0^\infty dk f_0^\nu(k).
\eeq
Equation (\ref{eq:hch}) becomes, 
\beq
\vec{h}_{ch} = (\frac{-1}{3 \rho_n \bar{\sigma}_{ch}k^2 }) 
\{
1+ \alpha_1^{ch}\frac{k}{M} + \alpha_2^{ch} \frac{T}{M} 
\}
\vec\grad{f^{\nu}_0(k)},   \label{eq:hch2}
\eeq
with
\beq
\bar{\sigma}_{ch} = \frac{7Y_e}{3.5 \mp 1.5(1-2Y_e)}
 \frac{4 G_F^2 cos^2 \theta_c}{\pi}(3c_a^2 + c_v^2) 
\label{eq:hsig}
\eeq
Equations (\ref{eq:hch2}) and (\ref{eq:hsig}) are valid for both $\nu_e$
and $\bar\nu_e$.  At $Y_e = 0.5 $, $\bar{\sigma}_{ch}$ is the same for
reactions (\ref{eq:ch1}) and (\ref{eq:ch2}). 

The final step is to combine charged and neutral currents in
Eqs. (\ref{eq:hx}) and (\ref{eq:hch2}) 
to get the total current for $\nu_e$,
\beq
\vec{h}_{\nu_e} = -
\frac{f_0(1- f_0)}
     {3\rho_n \sigma_0 (1 \!+\! r_0) (1 \!+\! r) k^2  } 
&\Bigl\{& 
\frac{k}{T^2}
(1 + \tilde{\alpha}_1 \frac{k}{M}
   + \tilde{\alpha}_2 \frac{T}{M} )
\vec\grad{T}\nonumber \\
&\pm& 
(1 + \tilde{\beta}_1\frac{k}{M}
   + \tilde{\beta}_2\frac{T}{M} ) 
\vec\grad{\eta}
\Bigr\}.  \label{eq:he}
\eeq
Where 
\beq 
r  = \frac{   \bar{\sigma}_{ch} }{ \sigma_{0} (1+r_0)  },
\eeq
describes the relative size of the charged and neutral currents and
\beq
\tilde{\alpha_i}={\bar{\alpha_i}+ r \alpha_i^{ch}\over 1+r},\ \ \ \ \
\tilde{\beta_i}={\bar\beta_i + r \alpha_i^{ch}\over 1 +r},
\eeq
are averages of the charged and neutral current coefficients.
Note, $\alpha_i^{ch}=\beta_i^{ch}$.

Using (\ref{eq:he}), one can calculate the number $J_e$ and energy $F_e$ 
currents for $\nu_e$. The expressions are the same as in neutral case, 
but with different coefficients,
\beq
\vec{J_e} = -\frac{T}{6 \pi^2\rho_n \sigma_0(1+r_0)(1+r)}
          \Bigl\{ 
\bigl[ L_1(\pm \eta) [1+ \tilde\alpha_2 \frac{T}{M}]
                  + L_2(\pm \eta) \tilde\alpha_1 \frac{T}{M} \bigr]
\frac{\vec\grad T}{T}\nonumber \\
\ \ \ \ \ \ \  \pm \bigl[ L_0(\pm \eta) [1+ \tilde\beta_2  \frac{T}{M}]
                  + L_1(\pm \eta) \tilde\beta_1  \frac{T}{M} \bigr]
                  \vec\grad\eta 
\Bigr\},
\label{eq:Jch}
\eeq
\beq
\vec{F_e} = -\frac{T^2}{6 \pi^2 \rho_n \sigma_0(1+r_0)(1+r)}
\Bigl\{ \bigl[ L_2(\pm \eta) [1+ \tilde\alpha_2 \frac{T}{M}]
                  + L_3(\pm \eta)  \tilde\alpha_1 \frac{T}{M} \bigr]
\frac{\vec\grad T}{T}\nonumber \\
\ \ \ \ \ \ \ 
\pm \bigl[ L_1(\pm \eta) [ 1+ \tilde\beta_2  \frac{T}{M} ]
                  + L_2(\pm \eta)  \tilde\beta_1  \frac{T}{M} \bigr]
                  \vec\grad{\eta} 
\Bigr\}. 
\label{eq:Fch}
\eeq

Equations (\ref{eq:Jch},\ref{eq:Fch}) are major results of the present paper.
Inside the neutrino sphere diffusion is a good approximation to the 
neutrino transport.  We have calculated the first order recoil corrections
of order $T/M$ to the conventional lowest order diffusion results.  The
bared coefficients $\bar\alpha_i$, $\bar\beta_i$ are averages of neutral 
current scattering from neutrons and protons and are appropriate for 
$\nu_\mu$ and $\nu_\tau$.  Finally, the tilde coefficients $\tilde\alpha_i$
and $\tilde\beta_i$ are averages of the charged and neutral current 
coefficients and are appropriate for $\nu_e$.

We illustrate our results for the simple case of small neutrino chemical
potentials $\eta\approx 0$.  This is expected to be a reasonable approximation
a few seconds after collapse when many neutrinos have escaped.
We also assume $\vec\grad\eta\approx 0$.

For mu and tau neutrinos the currents are now,
\beq
\vec{J} = \vec{J_0} \Bigl\{ 1 + [\bar\alpha_2 + {L_2(0)\over L_1(0)}
\bar\alpha_1]{T\over M}\Bigr\}, \label{eq:Jx}
\eeq
\beq
\vec{F} = \vec{F_0} \Bigl\{ 1 + [\bar\alpha_2 + {L_3(0)\over L_2(0)}\bar
\alpha_1]{T\over M}\Bigr\},     \label{eq:Fx}
\eeq
where the lowest order currents are,
\beq
\vec{J_0}=-{L_1(0)\over 6\pi^2\rho_n \sigma_0 (1+r_0)}\vec\grad{T},
\label{eq:Jx0}  \\
\vec{F_0}=-{T L_2(0)\over 6\pi^2\rho_n \sigma_0(1+r_0)}\vec\grad{T},
\label{eq:Fx0}
\eeq
and the Fermi integrals $L_i(0)$ are listed in table II.
Most supernova simulations ignore all recoil corrections so the
currents are just $J_0$ and $F_0$.  For $\nu_e$ and $\bar\nu_e$ the
currents are,
\beq
\vec{J_e} ={\vec{J_0}\over 1+r} \Bigl\{ 1 + [\tilde\alpha_2 + 
{L_2(0)\over L_1(0)}
\tilde\alpha_1]{T\over M}\Bigr\}, \label{eq:Je}\\
\vec{F_e} = {\vec{F_0}\over 1+r} \Bigl\{ 1 + [\tilde\alpha_2 + 
{L_3(0)\over L_2(0)}\tilde\alpha_1]{T\over M}\Bigr\}. \label{eq:Fe}
\eeq

\begin{table}
\centering Table II. Integrals $L_i(\eta=0)$.  

\medskip
\begin{tabular}{cc}         \hline\hline \\
$i$                      &  $L_i(0)$      \\ \hline 
0  &  0.5    \\ 
1  &  ln2$=0.69314$   \\
2  &  $1.64493$     \\ 
3  &  $5.40926$  \\
\hline \hline
\end{tabular}
\end{table}

Figure 1 shows the energy current $F$ versus temperature $T$ for $\nu_{\mu}$,
$\nu_{\tau}$ (lower dotted line), $\bar\nu_{\mu}$, $\bar\nu_{\tau}$ (upper
dotted line), $\nu_e$ (lower solid line) and $\bar\nu_e$ (upper
solid line).  In the absence of $T/M$ recoil corrections all lines
in figure 1 would be horizontal.  Thus the slopes come from the
recoil corrections.  The ratio of neutrino currents to $F_0$
decreases with $T$ and the ratio of anti-neutrino currents to $F_0$ 
increases with $T$.  The different panels in fig. 1 are for different 
electron fractions $Y_e$. 

Fig. 2 shows the number current $J$.  This follows the same
trends as the energy current $F$.  However, the slope of $J/J_0$ versus $T$
is slightly smaller than for $F/F_0$ because the corrections grow with
$k/M$ and $F$ weighs high $k$ neutrinos more than $J$ does.
  Below, we focus on the first panels  $Y_e=0.1$ since at late times the 
star is very neutron rich.

All of these currents have the approximate form,
$\vec C \approx {\vec C\,}^0 ( 1 \mp 10 {T\over M})$,
where $C^0$ is a generic neutrino current (for example $F_e$) without
$T/M$ corrections.  Slightly more accurate coefficients $\lambda_i$
for,
\beq
\vec C={\vec C\,}^0(1+\lambda_i {T\over M}),
\label{eq:lambda_i}
\eeq 
and $Y_e=0.1$ are collected in Table III.

\begin{table}
\centering Table III. Coefficients $\lambda_i$ of $T/M$ corrections.  
The minus sign is for neutrinos and the plus sign for anti-neutrinos.
\medskip

\begin{tabular}{cc}         \hline\hline \\
Current                      &  $\lambda_i$       \\ \hline 
$J_e$  &  $1.06\mp  8.76$    \\ 
$F_e$  &  $4.14\mp 12.14$    \\
$J_x$  &  $1.85\mp  7.75$    \\ 
$F_x$  &  $5.25\mp 10.74$    \\
\hline \hline
\end{tabular}
\end{table}

The currents are proportional to neutrino mean free paths, assuming
one is well inside the neutrino sphere.  Most previous literature claimed
the mean free paths $\lambda_i$ are related as follows,
\beq
\lambda_{\nu_e} < \lambda_{\bar\nu_e} < \lambda_{\nu_x} \approx 
\lambda_{\bar\nu_x},
\label{eq:normal}
\eeq
with $x=\mu$ or $\tau$.  This is because $\nu_e$ and $\bar\nu_e$ have
additional charged current interactions compared to $\nu_x$.  Furthermore,
$\lambda_{\nu_e}$ is less than $\lambda_{\bar\nu_e}$ because there
are more neutrons than protons for the neutrinos to capture on. 
Indeed Fig. 1 agrees with Eq. (\ref{eq:normal}) for low temperatures.

However, at high temperatures this picture changes.  Above $T \approx 20$ MeV
(for $Y_e=0.1$) the order becomes,
\beq
\lambda_{\nu_e} < \lambda_{\nu_x} < \lambda_{\bar\nu_e} < \lambda_{\bar\nu_x}.
\label{eq:invert}
\eeq
Now all flavors of neutrinos have shorter mean free paths than all
flavors of anti-neutrinos.  This is because neutrino cross sections are 
intrinsicly larger than anti-neutrino cross sections.  Furthermore, this 
difference in cross section more than compensates for the extra charged 
current interactions of the $\bar\nu_e$.  {\it Thus at high $T$, all neutrinos 
have shorter mean free paths than all anti-neutrinos.}

Note, Figs. 1-2 assume zero chemical potentials.
In a full simulation the difference between neutrino and anti-neutrino
currents will lead to nonzero neutrino chemical potentials.  This is true 
even for $\mu$ and $\tau$ neutrinos.   Indeed, ref. \cite{munumber} discusses 
how the system quickly reaches a steady state equilibrium where the $\nu_x$
chemical potentials rise until the extra neutrino density balances the larger
anti-neutrino mean free path.  In this steady state equilibrium the
neutrino and anti-neutrino currents are once again equal.  However, there is
then a large excess of $\nu_x$ over $\bar\nu_x$ in the protoneutron star.
Full simulations should be run to further explore this feed back of
the different $\nu$ and $\bar\nu$ currents on the neutrino chemical potentials.

Finally, we consider diffusion of electron neutrinos at early
times when the chemical potential $\mu_{\nu_e}$ is large. 
In this case, the $\bar\nu_e$ current is very small.
We assume $\vec\grad T\ll \vec\grad \mu$.  The
ratio of Fermi integrals for large $\eta$ is just 
$L_{i+1}(\eta) / L_i(\eta)\approx \eta$ so that for $\nu_e$ only,
\beq
\vec J_e = \vec {J_e\,}^0 (1 +\tilde\beta_2  {T\over M} + \tilde\beta_1 
{\mu_{\nu_e} \over M} ),
\eeq
\beq
\vec F_e = \vec {F_e\,}^0 (1 +\tilde\beta_2  {T\over M} + \tilde\beta_1 
{\mu_{\nu_e} \over M} ).
\eeq
In principle, the coefficient $\tilde\beta_1$ is important 
because $\mu_{\nu_e}$ can be large $\approx 250$ MeV.  
However,  there is some cancellation between the 
$\epsilon_i$ terms in Eq. (\ref{eq:epsilon}) and the $\delta$ term of
Eq. (\ref{eq:delta}).   As a result $\tilde\beta_1$ is small.  
At $r_0\approx 0.1$ and  $r\approx 1$ we have $\tilde\beta_1\approx 0.5$. 
The coefficient $\tilde\beta_2\approx -6.6$ can make a significant 
contribution.  At a temperature of 50 MeV it reduces currents by 
$\approx 35\%$.

We now summarize and conclude.  We are interested in corrections to neutrino
currents in supernovae from the intrinsic differences between neutrino-nucleon
and anti-neutrino-nucleon interactions because of $C$ violation in the 
Standard  Model.  Charge conjugation violation shows up at recoil order $k/M$
 where the  neutrino energy $k$ is of order the temperature $T$.  
This is because time reversal symmetry forbids $\nu-N$ and $\bar\nu-N$ 
differences if the nucleon does not recoil.

In this paper we have calculated corrections
of order $T/M$ to neutrino transport inside a protoneutron star.
We expanded neutrino-nucleon cross sections to order $k/M$.  We then 
calculated appropriate angular moments of the Boltzmann equation.

There are three kinds of $T/M$ corrections.  The first involves the kinematics
of the nucleon recoiling after it is struck by the neutrino.  This is
described by the $\epsilon_i$ terms in Eq. (\ref{eq:epsilon}) 
and increases both neutrino and anti-neutrino mean free paths.  

The second correction arises from the thermal motion of the struck nucleons.
Note, the first order $p/M$ Doppler shift,  
with $p$ the nucleon momentum, averages to zero.  However, 
the second order $p^2/M^2\propto T/M$ contribution is nonzero and
decreases both neutrino and anti-neutrino mean free paths.  This gives rise to
the $\zeta_i$ terms in Eq. (\ref{eq:zeta}).  We find considerable, but not 
perfect, cancellations between theses two corrections 
($\epsilon_i$ and $\zeta_i$).

The final correction involves the $C$ violating nature of the weak 
interactions. The $\delta$ term in Eq. (\ref{eq:delta}) increases the 
anti-neutrino and decreases the neutrino mean free path.  
This term would be zero if the weak interactions
conserved $C$.  The $\delta$ coefficient makes a significant contribution 
because it involves the large $F_2$ or weak magnetism term.
The cancellation between recoil and thermal motion effects leaves the total
$T/M$ correction dominated by the $\delta$ term and of opposite sign for 
neutrinos compared to anti-neutrinos.

The $T/M$ corrections are large and can even change the relative size
of $\bar\nu_e$ and $\nu_x$ mean free paths.  These corrections should be 
important for quantities sensitive to differences between neutrinos and
anti-neutrinos.  For example, the large mu and tau numbers of supernovae
arise because of these $T/M$ corrections\cite{munumber}.
In addition, the ratio of neutrons to protons in the neutrino
driven wind above a proto-neutron star is sensitive to small
differences between neutrinos and anti-neutrinos and this may have large
implications for r-process nucleosynthesis\cite{H+L3}.

The impact of these $T/M$ corrections on large scale properties of a 
supernovae may allow one to identify macroscopic manifestations of the 
$C$ violation in microscopic weak interactions.  We believe it is 
important to study this further because supernovae are perhaps the
only present day large scale systems dominated by the weak interaction.  
Future work should include these $C$ violating corrections in detailed 
supernova simulations\cite{H+L5}.

This work is supported in part by U.S.Department of Energy under 
Contract No. DE-FG2-87ER40365 .

\begin{figure}[ht]
\centering
\includegraphics[width=4.0in,height=3.0in]{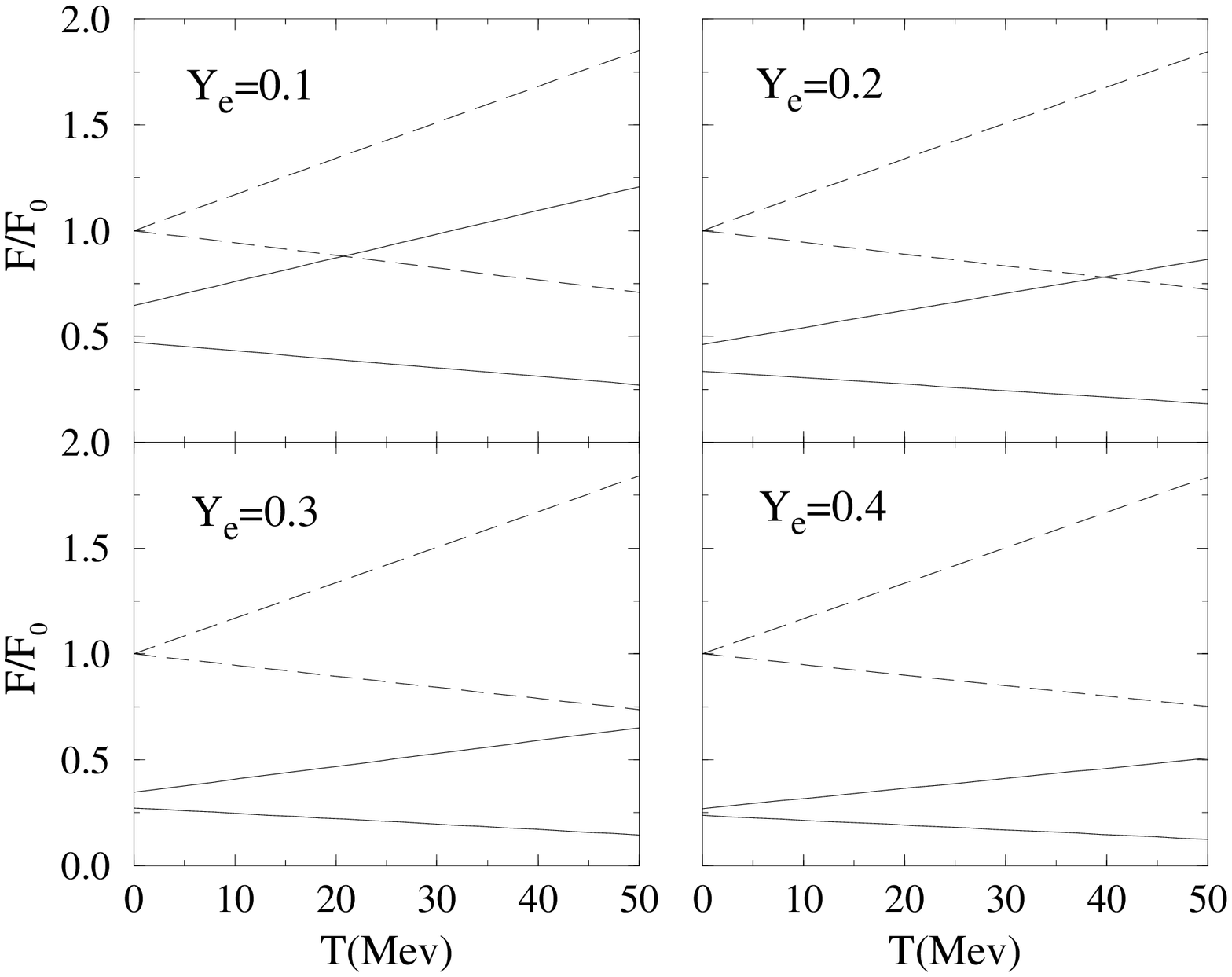}
\caption{ \small {
Neutrino energy current $F$ divided by $F_0$ versus temperature $T$.
Solid lines are for $\nu_e$ (lower) and $\bar{\nu_e}$ (upper curve), see
Eq. (\ref{eq:Fe}). 
Dashed lines are for  $\nu_x$ (lower) and $\bar{\nu}_x$ (upper) 
with $x=\mu$ or $\tau$, see Eq. (\ref{eq:Fx}).
The $\nu_x$ current without any $T/M$ corrections is $F_0$, 
see Eq. (\ref{eq:Fx0}).   
The four panels are for the indicated values of the electron fraction $Y_e$.
}
}
\end{figure}

\begin{figure}[!ht]
\centerline{\epsfig{file=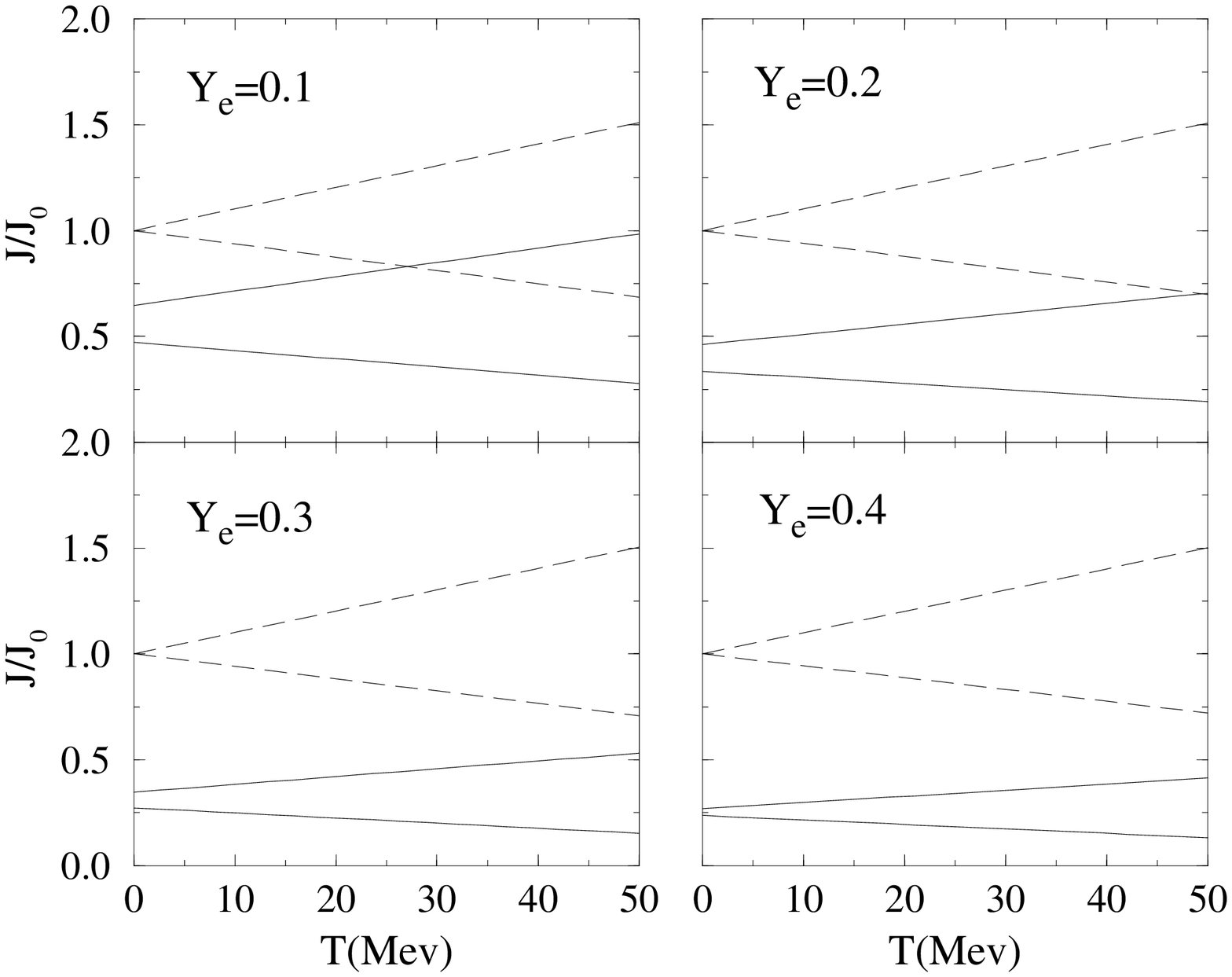,width=4.0in,height=3.0in}}
\caption{ 
Neutrino number current $J$, see Eqs. (\ref{eq:Je}, \ref{eq:Jx}), 
divided by $J_0$, Eq. (\ref{eq:Jx0}).  
The curves are labeled as in Fig. 1.
}
\end{figure}


\begin{thebibliography}{99}
\bibitem{parity} A. Vilenkin, 1979 unpublished; ApJ. {\bf 451} (1995) 700;
N.N. Chugai, Sov. Astron. Lett. {\bf 10} (1984) 87; 
C. J. Horowitz and J. Piekarewicz, Nuc. Phys. {\bf A640} 
(1998) 281;  C.J. Horowitz and Gang Li, Phys. Rev. Lett. {\bf 80} (1998)
3694,  Erratum-ibid. {\bf 81} (1998) 1985.

\bibitem{T+S} D.L.Tubbs and D. N. Schramm,Ap J {\bf 201}(1975) 467. 
\bibitem{B+P} B.T. Goodwin and C.J.Pethick, Ap J {\bf 253}(1982) 816.
\bibitem{S+S} R.F.Sawyer and A.Soni, Ap J {\bf 230}(1979) 859.
\bibitem{B+V} S. Bludman and K. VanRiper, Ap J {\bf 224}(1978) 631.
\bibitem{Iwamoto}N. Iwamoto (1981) Ph.D. thesis, University of Illinois,Urbana.
\bibitem{Schinder} P.J.Schinder, Ap J Suppl. {\bf 74}(1990) 249.

\bibitem{B+L} A. Burrows and J.M.Lattimer, ApJ {\bf 307} (1986) 178. 
\bibitem{K+J} W.Keil and H.-Th. Janka, A\&A {\bf 296} (1995)145.
\bibitem{munumber} C.J.Horowitz and Gang Li, Phys. Lett. {\bf B443} (1998)58.

\bibitem{H+L3} C.J. Horowitz and Gang Li, Phys. Rev. Lett. {\bf 82} (1999) 5198.
\bibitem{H+L5} C.J. Horowitz and Gang Li,  In preparation.   


\bibitem{H+W} C.J. Horowitz and K. Wehrberger, Phys. Rev. Lett. {\bf 66}(1991) 
271. 
\bibitem{Reddy} S. Reddy, M. Prakash and J.M. Lattimer, Phys. Rev. {\bf D58}
(1998)013009.
\bibitem{J+M}  H.-Th. Janka, E. M\"uller, ApJ Lett. {\bf 109}(1995) 448.



\bibitem{W+S+G} S. Weinberg, Phys. Rev. Lett. {\bf 19}(1967) 1264, 
A. Salam, Elementary Particle Theory, (ed. N. Svartholm), 
Almquist and Wiksells, Stockholm, 1968, 
S. L. Glashow, Nucl. Phys. {\bf 22}(1961) 579.

\bibitem{I+N}V.S.Imshennik and D.K.Nadezhin, Sov.Phys. JETP {\bf 36} (1973)
 821.

 

\end{thebibliography}
\end{document}